\documentclass[aps,eqsecnum,prd,twocolumn]{revtex4}
\usepackage{graphics,graphicx}
\usepackage{amsmath}
\usepackage{amssymb,latexsym,mathrsfs}
\usepackage{hyperref}

\def\bea{\begin{eqnarray}}
\def\eea{\end{eqnarray}}
\def\ba{\begin{array}}
\def\ea{\end{array}}

\def\beq{\begin{equation}}
\def\eeq{\end{equation}}

\begin{document}

\title{Hartman effect and dissipative quantum systems}

\author{Samyadeb Bhattacharya$^{1}$ \footnote{sbh.phys@gmail.com}, Sisir Roy$^{2} $ \footnote{sisir@isical.ac.in}}
\affiliation{$^{1,2}$Physics and Applied Mathematics Unit, Indian Statistical Institute, 203 B.T. Road, Kolkata 700 108, India \\}

\vspace{2cm}
\begin{abstract}

\vspace{1cm}

\noindent The dwell time for dissipative quantum system is shown to increase with barrier width. It clearly precludes Hartman effect for dissipative systems. Here calculation has been done for inverted parabolic potential barrier.
\vspace{2cm}

\textbf{ PACS numbers:} 03.65.Xp, 03.65.Yz \\

\vspace{1cm}
\textbf{Keywords:} Hartman effect, Dwell time, Dissipative system.

\end{abstract}

\vspace{1cm}

\maketitle

\section{Introduction}
In the recent years Hartman's work on tunneling time has drawn much attention of the community \cite{15}. At the time of it's publication, little attention was paid to Hartman's theoretical work on tunneling time of wavepackets in the sixties. Hartman analyzed the temporal aspects of tunneling by writing down solutions of the time-dependent Schr\"{o}dinger equation in terms of a superposition integral over stationary state solutions weighted by a Gaussian momentum distribution function. Without explicitly evaluating the integrals he could infer certain properties of the transmitted wave packet by examining the magnitude and phase of the integrand.  His main striking result was that under certain circumstances (opaque barrier) the tunneling time is independent of barrier length and the traversal time can be less than the time that would be required to travel a distance equal to the barrier length in vacuum. Many physicists hesitated to deal with Hartman's results since a very fast tunneling, or a zero tunneling time holds a serious consequence of violating Einstein's postulate of Special Theory of Relativity. Hartman effect was first re-examined by Fletcher \cite{15a} within stationary phase method for quasi-monochromatic non-relativistic particles tunneling through the potential barriers. Furthermore, on recent times, experiments with photonic band gap structures \cite{16,17} showed apparent superluminality. These observations as well as the theoretical predictions lead towards the phenomena of superluminal barrier traversal. Regarding the explanation of this apparent superluminality, it is important to mention some publications over the past two decades or so \cite{8a,8b,9,10,11,12,13,14}.  Some suggestions have been made \cite{9,10,11,12,13} to explain faster than light phenomena by the concept of energy storage and release in the barrier region. The argument is that the group delay, which is directly related to the dwell time by an additive self interference term \cite{14}, is actually the lifetime of stored energy (or stored particles) leaking through both ends of the barrier.
The relation between group delay ($\tau^G$), dwell time ($\tau^D$) and self-interference delay ($\tau^I$) is given by
\beq\label{1.1}
\tau^G=\tau^D+\tau^I
\eeq
When the reflectivity is high, the incident pulse spends much of it's time dwelling in front of the barrier as it interferes with itself during tunneling process. This excess dwelling is interpreted as the self-interference delay. This term can be successfully disentangled from the dwell time \cite{14}. Now if the surrounding of the barrier is dispersionless, then the self-interference term vanishes, resulting in the equality of group delay and dwell time \cite{12}. In that case the dwell time will give the lifetime of energy storage in the barrier region. In a relatively recent work \cite{17a}, Jakiel et.al. have shown that Hartman effect is valid for all known expressions of the mean tunneling time, in various non-relativistic approaches, with finite width barriers without absorbtion or dissipation. In a recent paper \cite{17b} we have formulated the expression of dwell time in presence of dissipation by the formalism of weak measurement. We have shown that inclusion of dissipative interaction precludes the zero time tunneling. But in that work we have not discussed the behavior of dwell time with increasing barrier thickness. In this paper, it is our aim to calculate the dwell time theoretically for tunneling through a dissipative inverted parabolic barrier to find whether ``Hartman effect" exists in presence of dissipative interaction with the environmental bath modes; i.e. whether the dwell time saturates with increasing barrier thickness or not. The inverted harmonic potential or parabolic barrier, which provides an infinite potential barrier can be used as a toy model for tunneling, where exact Gaussian wave packets may be found as solutions. The inverted harmonic oscillator problem attracts a great deal of attention not only for being one of the exactly solvable potential in quantum mechanics \cite{18,19,20,21,22,23,24,25} but also because of having wide range of applicability in many branches of physics. It receives quite a formidable number of applications in many branches of physics from high energy  to solid state theory. To incorporate dissipation, we will use the formalism of Yu \cite{26} and Yu et.al. \cite{27}, where evolution of wave function and tunneling in dissipative system has been discussed. In Section II we will briefly discuss the background formalism, in Section III we will derive the dwell time using the discussed formalism and finally in Section IV we will summarize with some concluding remarks.

\section{Background formalism}

Yu et.al. \cite{27} have shown that for the simplest example of dissipative system, a harmonic oscillator coupled to a heat bath in a special case of ohmic dissipation, a self contained treatment can be used to get wave function evolution, where path integral technology need not be used. Later in another paper \cite{26}, Yu used this particular formalism in the quantum tunneling case for a inverted harmonic oscillator to obtain the tunneling probability and current density. Here we will use this formalism to get the dwell time for such systems. Using the notations of Ref. \cite{26,27} and quoting formulae from there, we first identify the total Hamiltonian of the system of inverted harmonic potential with a coupled infinite harmonic oscillator heat bath as
\beq\label{2.1}
H= \begin{array}{ll}
    \frac{P^2}{2M}-\frac{1}{2}M(\omega_0^2+\Delta\omega^2)q^2+ q\sum_j c_j x_j \\
     + \sum_j \left(\frac{p_j^2}{2m_j}+\frac{1}{2}m_j\omega_j^2x_j^2\right)
   \end{array}
\eeq
From this Hamiltonian, the quantum Langevin equation can be derived as
\beq\label{2.2}
\ddot{q}+\eta \dot{q}-\omega_0^2 q=f(t)
\eeq
where $\eta$ is the damping constant and $f(t)$ is the Brownian motion driving force given by
\beq\label{2.3}
f(t)= -\sum_j \frac{c_j}{M}\left(x_{j0}\cos \omega_j t + \dot{x}_{j0} \frac{\sin \omega_j t}{\omega_j}\right)
\eeq
This equation (\ref{2.2}) can be solved to give
\beq\label{2.4}
q(t)= a_1(t)q_0 + a_2(t)\dot{q}_0 + \sum_j \left[ b_{j1}x_{j0}+ b_{j2}\dot{x}_{j0} \right]
\eeq
where $q_0,\dot{q}_0,x_{j0},\dot{x}_{j0}$ are the initial position and velocity of the system and the bath respectively, and
\beq\label{2.5}
\begin{array}{ll}
a_1=e^{-(\eta/2)t}\left(\cosh \omega t + \frac{\eta}{2\omega}\sinh \omega t\right),\\
a_2=e^{-(\eta/2)t}\frac{\sinh \omega t}{\omega}
\end{array}
\eeq
\beq\label{2.6}
\begin{array}{ll}
b_{j1}= -\frac{c_j}{M}\int_0^t a_2(t')\cos\omega(t-t')dt',\\
b_{j2}= -\frac{c_j}{M}\int_0^t a_2(t')\sin\omega(t-t')dt'
\end{array}
\eeq
with $\omega=(\omega_0^2+\eta^2/4)^2$.
The solution can be used to show that the wavefunction of the system plus bath can be written as \cite{27}
\beq\label{2.7}
\psi (q,\{\xi_j\},t)= \psi\left(q-\sum_j\xi_j,t\right) \mathbf{\Pi}_j ~\chi_j(\xi_j,t)
\eeq
Now let us assume that the initial wavefunction is a Gaussian wave packet centered at the right of the peak of the potential by $z_0$ has the form
\beq\label{2.8}
\psi_0(q,t=0)=(2\pi\sigma^2)^{-1/4}e^{-(q-z_0)^2/4\sigma^2+ikq}
\eeq
Following \cite{27}, the Green's function can be derived as
\beq\label{2.9}
G(q,q_0;t,0)=\begin{array}{ll}
             \left(\frac{M}{2\pi i\hbar a_2}\right)^{1/2}\\
             \times \exp\left[\frac{iM}{2\hbar a_2}(a_1 q_0^2 + a_2 e^{\eta t}q^2 -2q_0 q)\right]
             \end{array}
\eeq
Then the wavefunction $\psi(q,t)$ can be calculated as
\beq\label{2.10}
\psi(q,t)=\begin{array}{ll}
          (2\pi\sigma^2)^{-1/4} (a_1+i\omega_0 a_2 r^2)\\
           \times e^{-(q-z_0)^2/4\sigma^2+i(c_2 q^2 + c_1 q + c_0)}
           \end{array}
\eeq
This is a Gaussian distribution with a width of
\beq\label{2.11}
\sigma_{\theta}^2=\sigma^2 (a_1^2+r^4\omega_0^2 a_2^2)
\eeq
where $r=\sigma_0/\sigma$ and $q_c=a_1 z_0 + a_2\hbar k/M$. The coefficients in the exponential term are
\beq\label{2.12}
c_2=\frac{Me^{\eta t}}{4\hbar}\frac{d}{dt}(\ln \sigma_{\theta}^2)
\eeq
\beq\label{2.13}
c_1=\frac{Me^{\eta t}}{4\hbar} \left(-2q_c\frac{d}{dt}(\ln \sigma_{\theta}^2)+4\dot{q}_c\right)
\eeq
\beq\label{2.14}
c_0=\frac{ka_2}{4}e^{\eta t}\left(q_c\frac{d}{dt}(\ln \sigma_{\theta}^2)-2\dot{q}_c\right)+\frac{q_c z_0}{4\sigma_{\theta}^2}\omega_0 a_2 r^2
\eeq
The solution equation (\ref{2.10}) of the time dependent Schr\"{o}dinger equation should satisfy the expression for the current density
\beq\label{2.15}
J=\frac{\hbar}{2Me^{\eta t}}\left(\psi \frac{\partial \psi^{*}}{\partial q}-\psi^{*} \frac{\partial \psi}{\partial q}\right)=\frac{\hbar}{Me^{\eta t}}|\psi|^2 (2c_2 q+c_1)
\eeq
Putting the value of $\psi,c_1,c_2$ we get
\beq\label{2.16}
J=\left(2(q-q_c)\frac{d}{dt}(\ln \sigma_{\theta}^2)+4\dot{q}_c\right)\times \frac{(a_1^2+r^4\omega_0^2 a_2^2)}{\sqrt{2\pi \sigma^2}}e^{-\frac{(q-z_0)^2}{2\sigma^2}}
\eeq
We will utilize equation (\ref{2.16}) to get the dwell time in the next section.

\section{Dwell time in dissipative medium and Hartman effect}
Dwell time is defined as the average number of particles within the barrier region divided by the average number entering (or leaving) the barrier per unit time. It corresponds to the average time spent by a particle within the barrier irrespectively of whether it is finally reflected or transmitted. If we follow Winful's explanation \cite{12}, when the surroundings of the barrier is dispersionless, it represents the lifetime of energy storage in the barrier region. The dwell time in a neighborhood of $q$ is defined as the ratio between the particle number in the interval $[q, q + dq]$ and the incoming current
\beq\label{3.1}
d\tau^D=\frac{|\psi(q)|^2}{J_{in}}
\eeq
Obviously, Eq. (\ref{3.1}) describes a balance equation: in the stationary case the injected current equals the decay
rate of the probability in $[q, q +dq]$. The dwell time $\tau^D$ of a finite region within the context of a stationary state scattering problem is obtained via a spatial integration of Eq. (\ref{3.1}). So the dwell time $\tau^D$ is given by
\beq\label{3.2}
\tau^D=\frac{M}{\hbar k}\int_{q_0}^q|\psi(q)|^2 dq= \frac{M}{\hbar k}\int_{q_0}^q \rho(q) dq
\eeq
where $J_{in}=\frac{\hbar k}{M}$ is the incident flux. We are considering constant incident flux for simplicity.
Now we are dealing with non-stationary quantum states interacting with the environment. So the wave function and consequently the probability density function will also be dependent on time. Differentiating with respect to time:
\beq\label{3.3}
\dot{\tau}^D= \frac{M}{\hbar k}\int_{q_0}^q \frac{\partial \rho(q,t)}{\partial t} dq
\eeq
Using the continuity equation:
\beq\label{3.4}
\dot{\tau}^D=-\frac{M}{\hbar k}\int_{q_0}^q \frac{\partial J(q,t)}{\partial q} dq=\frac{M}{\hbar k}[J(q_0)-J(q)]
\eeq
Now integrating equation (\ref{3.4}) with respect to time:
\beq\label{3.5}
\tau^D=\frac{M}{\hbar k}\int_0^T[J(q_0)-J(q)]dt
\eeq
where measurements are made at $t=0$ and $t=T$. For long time measurement, we can set $T=\infty$. As we have discussed that the initial wavefunction is centered at the right of the peak of the potential, it is travelling from right to left. So let us now set $q_0=2z_0$ and $q=0$. Then using equation (\ref{2.16}) we get
\beq\label{3.6}
J(q_0)-J(q)= 4z_0\frac{d}{dt}(\ln \sigma_{\theta}^2)\frac{(a_1^2+r^4\omega_0^2 a_2^2)}{\sqrt{2\pi \sigma^2}}e^{-\frac{z_0^2}{2\sigma^2}}
\eeq
Considering the spread of the wavefunction to be time dependent and replacing $\sigma$ be $\sigma_{\theta}$ in equation (\ref{3.6}), then putting in equation (\ref{3.5}), we get
\beq\label{3.7}
\tau^D~~~~~~\begin{array}{ll}
=\frac{M}{\hbar k}\int_0^{\infty} 4z_0\frac{(a_1^2+r^4\omega_0^2 a_2^2)}{\sqrt{2\pi \sigma_{\theta}^2}}e^{-\frac{z_0^2}{2\sigma_{\theta}^2}}\frac{d}{dt}(\ln \sigma_{\theta}^2)dt \\
=\frac{M}{\hbar k \sqrt{2\pi}}\left(\frac{4z_0}{\sigma}\right)\int_{t=0}^{\infty}\frac{\exp[-\frac{(z_0^2/2\sigma^2)}{x^2}]}{\sqrt{x^2}}d(x^2)

\end{array}
\eeq
where $x^2=a_1^2+r^4\omega_0^2 a_2^2$. Putting $\frac{z_0}{\sqrt{2}\sigma}=\zeta$:
\beq\label{3.8}
\tau^D=\frac{4M\zeta}{\hbar k \sqrt{\pi}} \int_{t=0}^{\infty}\frac{\exp(\frac{-\zeta^2}{x^2})}{\sqrt{x^2}}d(x^2)
\eeq
Again we substitute $y=\zeta/x$. Now as $t\rightarrow 0, x\rightarrow 1 ~~~~\mbox{and}~~~~ y\rightarrow \zeta$; as $t\rightarrow\infty, x\rightarrow\infty~~~~ \mbox{and}~~~~ y\rightarrow 0$. So equation (\ref{3.8}) becomes
\beq\label{3.9}
\tau^D= \frac{8M\zeta}{\hbar k \sqrt{\pi}}\int_{y=0}^{\zeta} \frac{e^{-y^2}}{y^2}dy
\eeq
The remaing problem is that the integration do not converge at $y=0$. So instead of taking the limit of $x$ to be $\infty$, we take a finitely long time $t=T_{long}$, for which $x\rightarrow \zeta$ and $y\rightarrow 1$.
So now the dwell time is found to be
\beq\label{3.10}
\tau^D~~~~~~\begin{array}{ll}
            = \frac{8M\zeta}{\hbar k \sqrt{\pi}} \int_{1}^{\zeta} \frac{e^{-y^2}}{y^2}dy \\
            = \frac{8M\zeta}{\hbar k \sqrt{\pi}} \left[\frac{1}{e}-\frac{e^{-\zeta^2}}{\zeta} + Erf(1)-Erf(\zeta)\right]
\end{array}
\eeq
Here `$Erf$' stands for error function. Generally, the width of the potential ($z_0$) is greater than the width of the wavefunction ($\sigma$). Also we will study the behavior of dwell time with increasing width of the barrier. Therefore the exponentially decaying term $e^{-\zeta^2/}\zeta$ would be less  significant. So neglecting this term from equation (\ref{3.10}) we get
\beq\label{3.11}
\tau^D=\frac{8M\zeta}{\hbar k \sqrt{\pi}} \left[\frac{1}{e}+ \sqrt{\pi}(Erf(1)-Erf(\zeta))\right]
\eeq
Expressing the dwell time in terms of potential width $w=2z_0$, we get
\beq\label{3.12}
\tau^D \begin{array}{ll}
        =\frac{2M}{\hbar k \sqrt{\pi}}\left(\frac{w}{\sigma}\right)^2 \left[\frac{1}{e}+ \sqrt{\pi}\left(Erf(1)-Erf(w/2\sigma)\right)\right]\\
        =\frac{2M}{\hbar k \sqrt{\pi}}F(\frac{w}{\sigma})
        \end{array}
\eeq
\begin{figure}[htb]
{\centerline{\includegraphics[width=7cm, height=4cm] {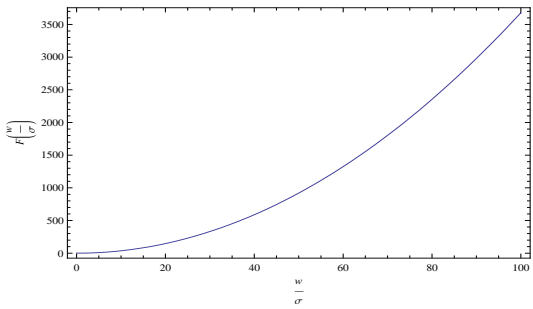}}}
\caption{$F(\frac{w}{\sigma})$ with the scaled potential width $\frac{w}{\sigma}$}
\label{figVr}
\end{figure}

The figure shows that in presence of dissipative interaction, the dwell time is not saturating with increasing barrier width, but it increases as the barrier width increase. Now let us compare this result with the classical case of a particle traveling in presence of a frictional force proportional to velocity ($v$) having the equation of motion
\beq\label{3.13}
\dot{v}+\gamma v=0
\eeq
The solution of this equation will be of the form
\beq\label{3.14}
v=v_0 e^{-\gamma t}
\eeq

Consider that the classical particle takes $\tau^{CL}$ time to travel the distance $w_{cl}$ in presence of the usual velocity dependent frictional force. Then integrating equation (\ref{3.14}), we get
\beq\label{3.15}
\tau^{CL}= \frac{1}{\gamma} \ln \left(\frac{1}{1-\frac{\gamma}{v_0}w_{cl}}\right)
\eeq
 A condition must be imposed that $\gamma \leq \frac{1}{\tau_0}$, where $\tau_0=\frac{w_{cl}}{v_0}$ is the time taken by the particle to cross the distance $w_{cl}$ with constant velocity $v_0$ in absence of dissipation. Because otherwise the particle will lose all of it's kinetic energy before crossing the barrier. Now considering $ \gamma \ll \frac{1}{\tau_0} $ and expanding the logarithmic term in equation (\ref{3.15}) and neglecting 3rd and higher order terms, we get
\beq\label{3.16}
\tau^{CL}= \alpha w_{cl}^2 + \beta w_{cl}
\eeq
where $\alpha=\gamma/v_0^2$ and $\beta=1/v_0$.
\begin{figure}[htb]
{\centerline{\includegraphics[width=7cm, height=4cm] {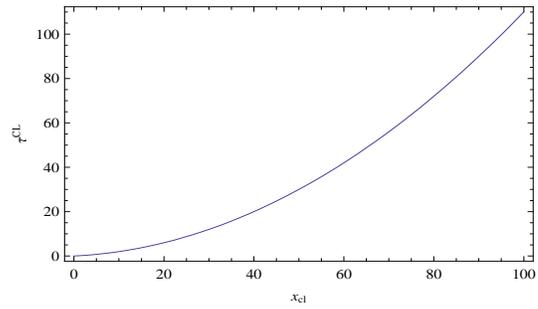}}}
\caption{$\tau^{CL}$ with $w_{cl}$ taking $\alpha=.01$ and $\beta=.1$}
\label{figVr}
\end{figure}
From FIG.1 and FIG.2 we see that the increment of dwell time with barrier width follow almost the same nature of that of a classical particle in presence of velocity dependent frictional force. So here we find that inclusion of dissipative interaction makes the behavior of dwell time quasi-classical. Also such dependence of dwell time on the barrier width precludes the afore mentioned ``Hartman effect".

\section{Conclusion}
It is evident from the above analysis that due to interaction with the environmental bath modes the tunneling time does not saturate with increment of barrier width. Unlike Hartman effect, in this case the dwell time depends on barrier thickness. As we have mentioned earlier, explanation of the saturation of tunneling time can be given by the concept of energy storage and release in the barrier region \cite{13}. Group delay, which is equal to dwell time in absence of self interference, is proportional to the stored energy and it saturates (in absence of dissipation), as the stored energy saturates. In presence of dissipative interaction, tunneling entity loses energy to the interacting bath modes and in the process the saturation of energy is prevented. As the barrier length increases, the interaction with the bath modes also increase, resulting more and more energy loss. So the dwell time increases with increase of barrier length instead of getting saturated. The continuous interaction with the environmental bath modes makes the behavior of dwell time  quasi-classical.

\end{document}